\documentclass[manuscript,reviewscreen,acmsmall,authorversion,nonacm]{acmart}
\pdfoutput=1

\AtBeginDocument{%
  \providecommand\BibTeX{{%
    \normalfont B\kern-0.5em{\scshape i\kern-0.25em b}\kern-0.8em\TeX}}}

\setcopyright{acmcopyright}
\copyrightyear{2018}
\acmYear{2018}
\acmDOI{10.1145/1122445.1122456}

\acmJournal{JACM}
\acmVolume{37}
\acmNumber{4}
\acmMonth{8}

\graphicspath{{figures/}}

\usepackage{lipsum}
\usepackage{longtable}

\begin{document}
%\begin{sloppypar}

\title[Understanding The Concerns and Choices of The Public When Using LLMs for Healthcare]{Understanding The Concerns and Choices of The Public When Using Large Language Models for Healthcare}

\author{Yunpeng Xiao}
\email{yxiao28@hawk.iit.edu}
\affiliation{
  \institution{Illinois Institute of Technology}
  \city{Chicago}
  \state{Illinois}
  \country{United States}
}

\author{Kyrie Zhixuan Zhou}
\email{zz78@illinois.edu}
\affiliation{
  \institution{University of Illinois at Urbana-Champaign}
  \city{Champaign}
  \state{Illinois}
  \country{United States}
}

\author{Yueqing Liang}
\email{yliang40@hawk.iit.edu}
\affiliation{
  \institution{Illinois Institute of Technology}
  \city{Chicago}
  \state{Illinois}
  \country{United States}
}

\author{Kai Shu}
\email{kai.shu@emory.edu}
\affiliation{
  \institution{Emory University}
  \city{Atlanta}
  \state{Georgia}
  \country{United States}
}

\renewcommand{\shortauthors}{Yunpeng Xiao, Kyrie Zhixuan Zhou, Yueqing Liang, and Kai Shu}

\begin{abstract}
Large language models (LLMs) have shown their potential in biomedical fields. However, how the public uses them for healthcare purposes such as medical Q\&A, self-diagnosis, and daily healthcare information seeking is under-investigated. This paper adopts a mixed-methods approach, including surveys (N=214) and interviews (N=17) to investigate how and why the public uses LLMs for healthcare. We found that participants generally believed LLMs as a healthcare tool have gained popularity, and are often used in combination with other information channels such as search engines and online health communities to optimize information quality. Based on the findings, we reflect on the ethical and effective use of LLMs for healthcare and propose future research directions.
\end{abstract}




\maketitle
\section{Introduction}
Large language models (LLMs) have been a topic of great interest in recent years, especially in the field of natural language processing (NLP). LLMs are a type of AI model designed to generate human-like text by analyzing vast amounts of data. LLMs are based on deep learning techniques and typically involve neural networks with many layers and a large number of parameters, allowing them to capture complex patterns in the data they are trained on \citep{zhao2023survey}. The primary goal of LLMs is to understand the structure, syntax, semantics, and context of natural language, so they can generate coherent and contextually appropriate responses. ChatGPT is the first and currently most popular LLM-powered chatbot: as of December 2023, ChatGPT had more than 180 million users \citep{ChatGPTusers}.

The powerful performance of LLMs has also attracted many academics to discuss their applications in the fields of biomedicine and public health. ChatGPT is effective in providing patients with information and support in a variety of scenarios, including mental health assessments, counseling, medication management, and patient education \citep{cascella2023evaluating}. A recent analysis explored how ChatGPT and other LLMs could enhance medical education, simplify clinical decision-making, and improve patient outcomes \cite{sallam2023utility}.

Even before LLMs existed, Internet-based healthcare information-seeking channels had gained popularity. Internet-based healthcare information-seeking channels include search engines \cite{de2014seeking, rice2006influences}, online communities \cite{chung2017personal,loh2023social}, ask-the-doctor platforms \cite{ding2020getting, hao2015development}, healthcare chatbots \cite{you2020self, nadarzynski2019acceptability, pereira2019using} and so on. Internet-based self-diagnosis and healthcare information-seeking provide people with convenient and fast medical services without waiting for appointments or queuing, and help people better understand their health status and improve their health awareness. 
% Internet self-diagnosis can further provide users with more choices, such as online consultation and online purchase of drugs. 
However, Internet-based medical services have such disadvantages as inaccurate information ~\citep{de2014seeking,johnson2022cancer}, potentially leading to misdiagnosis or delayed treatment.
% Firstly, the accuracy and reliability of internet self-diagnosis need to be improved. 

Now LLMs also provide similar services and potentially bring new risks and opportunities. LLMs' impact as a healthcare information provider on public health, doctor-patient relationships, and other aspects of people's personal health has not been well investigated yet. This is because current research on the ethical issues of the use of LLMs in healthcare mainly focuses on clinical applications and support of health professionals and researchers, and there are few studies on public health~\citep{haltaufderheide2024ethics}.
% These are issues worthy of exploration and research. However, there is currently insufficient understanding of the public’s motivations for using LLMs for healthcare information and the risks that may arise from such uses. 
To bridge this research gap, we aspire to answer the following research questions through surveys and interviews:
\par• \textbf{RQ1:} How the public obtain healthcare information? 
\par• \textbf{RQ2:} What are the public’s motivations for using LLMs to obtain healthcare information?
\par• \textbf{RQ3:} What are the public’s choices and concerns about using LLMs to obtain health information?

% In the survey, to investigate the majority of the public's use of LLMs and their value judgments when facing controversial issues, we designed a questionnaire containing 25 questions. This questionnaire includes three parts. The first part is about basic information such as age and race; the second part is about what tools they use to search for health information and what kind of health information they search for; the third part is about value judgments on some controversial issues, such as doctor-patient relationship and misinformation. 

Through the surveys, we took a broader perspective to find how the public accesses healthcare information and gained a preliminary understanding of the issues that arise when the public obtains health information.

To gain a deeper understanding of the public’s choices and concerns about using LLMs to obtain health information, we conducted semi-structured interviews with 17 participants who had used LLMs in this regard.
% to better confirm the public’s intentions and thoughts.
We found that: (1) people have learned to use LLMs to obtain health information effectively, and through cross-validation with other information channels; (2) compared with search engines, LLMs provide healthcare information maybe more accurately and conveniently; (3) people with medical knowledge are more likely to discover the ``hallucinations'' of using LLMs to obtain health information, while for the general public, LLMs provide sufficient information to help them make self-diagnosis; (4) it is controversial whether doctors should use LLMs to diagnose patients and less so when it comes to using LLMs to perform auxiliary work such as writing medical records. (5) In the digital niche, LLMs have both competitive and complementary relationships with search engines and have complementary relationships with online communities and online consultations.

Based on the results obtained, we discuss future research directions at the intersection of LLMs and public health. We also provide suggestions for the public to more effectively use LLMs to search for health information, and for doctors to use LLMs for medical purposes. 
% We also provide some suggestions for future research on LLMs in the field of public health.

% \begin{figure*}[H]
%      \centering    
%      \includegraphics[width=0.5\textwidth]{images/searchLLM.png}
%      \caption{Image of health retrieval using LLMs drawn using Copilot (Microsoft's large language model). This image is also used for our survey distribution.}
%      \label{fig:searchLLM.png}
% \end{figure*}
\section{Related Work}
In this section, we first introduce the concept of LLMs and their development in the healthcare field. Then we introduce how the public accesses healthcare information. Finally, we introduce the specificity of health information and the emergent issues when the public accesses healthcare information. 

\subsection{LLMs in The Healthcare Field}
LLMs are advanced AI systems capable of understanding and generating human-like text. They are called ``large'' because they are trained on vast amounts of data and have a significant number of parameters that help them understand the nuances of language \citep{zhao2023survey}. From its development history, LLMs have evolved from transformers \citep{vaswani2017attention} to the current user-facing models such as ChatGPT. Today's LLMs like ChatGPT~\citep{ray2023chatgpt} and LLaMA~\citep{touvron2023llama} are essentially decoder-only transformers \citep{yang2023harnessing}. ChatGPT is the most popular LLM -- as of December 2023, ChatGPT had more than 180 million users \citep{ChatGPTusers}. 

The powerful performance of LLMs has attracted the attention of researchers in the healthcare field. Most research focuses on fine-tuning LLMs to better suit medical tasks~\citep{singhal2023large, liu2023evaluating, delsoz2023use,yang2022large}. Some studies discuss the application of LLMs in the field of medical education \citep{sallam2023utility, safranek2023role}. Others have explored the challenges of practical clinical application of LLMs~\citep{harrer2023attention, singhal2023large, sezgin2023artificial}.  Med-PaLM2 is
Google’s medical domain LLMs is the first LLM that can achieve an “expert” level of performance on the MedQA dataset of US Medical Licensing Examination (USMLE2)-style questions, with an accuracy of over 85\%~\citep{singhal2023towards}. 

However, in the field of public healthcare, there are currently few relevant studies. Some studies provide suggestions from the perspective of experts on how the public can use LLMs to obtain healthcare information~\citep{biswas2023role}. However, they do not provide an in-depth understanding of the general public's perspectives, so the proposals given are often from a technical perspective and lack a sociological perspective. In 2023, a study examined the benefits and challenges of Carecall, an LLM-based conversational robot developed in Korea, for public health applications~\citep{jo2023understanding}. Carecall is aimed at socially isolated individuals, while popular LLMs such as ChatGPT can potentially help the mass public access healthcare information. So in this paper, we aim to explore how the public utilizes general-purpose LLMs for healthcare purposes. A recent study explored the public's intention to use ChatGPT for self-diagnosis~\citep{shahsavar2023user}, but the survey was relatively simple -- it only investigated whether the public was willing to use ChatGPT for self-diagnosis, and ignored many other related issues. Instead, we aspired to provide a nuanced picture of users' motivations, experiences, and challenges when using LLMs for healthcare.

\subsection{How the Public Access Healthcare Information}

% Because the public generally obtains health information ``spontaneously'' through some methods (such as using search engines), we will elaborate on this in the next section. 

The Internet has greatly changed the form of information transmission, and information can be transmitted at an unprecedented speed. Before 2010, the public mainly obtained healthcare information from the Internet through search engines \citep{berland2001health,spink2004study,eysenbach2003prevalence,gray2005health}. After 2010, the methods of obtaining health information gradually diversified. Although search engines were still the most popular channel \citep{de2014seeking}, the public also began to try different methods to obtain health information. Some people used social media such as Twitter~\citep{de2014seeking, scanfeld2010dissemination}, WeChat~\citep{zhang2017public, hong2021understanding} or Instagram~\citep{chung2017personal} to obtain or share health information. Due to the shortage of medical resources, some online consultation platforms have emerged -- patients communicate with doctors online through medical platforms ~\citep{ding2020getting, hao2015development, hao2017tale}. With the rise of artificial intelligence technology, some AI-driven medical/healthcare chatbots have emerged in recent years ~\citep{pereira2019using}, such as Ada ~\citep{you2020self}. Patients or the general public can enter questions into the chatbots and get answers. Overall, the ways for the public to obtain health information are becoming increasingly diverse, which is in line with the trend of ``information explosion'' in the information society. Diverse methods of obtaining health information can be likened to "ecological niches" \cite{zhu2014selecting} in the digital ecosystem \cite{milton2024seeking, mcluhan1994understanding}, that is, like online communities, there is a competitive and complementary relationship between different methods of obtaining information. Previous studies have shown that there is heterogeneity in the acquisition of health information by search engines and social media, that is, users may search for different types of health information through search engines and social media \cite{de2014seeking, kanthawala2016answers}, and the two methods are complementary.

LLMs demonstrate more powerful natural language interaction capabilities and have penetrated the medical field \cite{li2023chatgpt}. After LLMs join the ecological niche, which methods will compete with them and which will complement them remain under investigation.

\subsection{Concerns When Accessing Healthcare Information using LLMs}
In the information explosion process, one of the prevalent concerns is that misinformation also explodes \cite{zhou2019fake, shu2017fake}. 
% This performance is multifaceted. 
In the field of public health, misinformation is particularly rampant \cite{suarez2021prevalence}. Misinformation related to serious diseases (e.g., cancer-related misinformation~\citep{johnson2022cancer, chen2018nature, warner2022online}) has often been circulated on social media, accompanied by advertisements for high-priced health products that had no actual therapeutic effect. Since the beginning of the COVID-19 pandemic, conspiracy theories have emerged -- for example, rumor went that the COVID-19 virus was spread by 5G technology~\citep{ahmed2020covid}. A wide range of misinformation about the COVID-19 pandemic existed on social media~\citep{enders2020different, kouzy2020coronavirus, evanega2020coronavirus, cuan2020misinformation}. Even more so, during the COVID-19 pandemic, it is not just the general public who may be spreading misinformation, but even professional doctors. One study identified 52 American doctors spreading misinformation on at least 5 different social platforms during the pandemic, which may have impacted millions of people~\cite {sule2023communication}. 

Existing studies have pointed out that ChatGPT could provide accurate health information if appropriately prompted~\cite {mehnen2023chatgpt}. This proves the huge potential of large language models in disease diagnosis. However, considering that LLMs are known to generate ``hallucinations'' \citep{alkaissi2023artificial, rawte2023survey}, understanding how the public experiences ``hallucinations'' when using LLMs to access healthcare information is important. One study showed that in a Turing test, patients had difficulty distinguishing whether the advice was given by a real doctor or generated by ChatGPT~\citep{nov2023putting}. Many doctors and scholars have warned that ``AI cannot replace doctors~\citep{mehnen2023chatgpt, li2023exploring, cheng2023potential}.'' Some researchers thought that although ChatGPT could provide useful information and answer general questions related to diseases, it cannot replace the expertise, training, and experience of a qualified doctor~\cite {cheng2023potential}. There is even a report of life-threatening delays in diagnosis due to the use of Chatgpt~\cite {saenger2024delayed}.

The doctor-patient relationship has always been a controversial topic, which can be complicated by LLMs. As early as 2006, there was research on doctors using search engines to diagnose conditions~\citep{tang2006googling}. In 2011, a survey showed that nearly half of doctors used Google and Yahoo to diagnose patients \citep{Searchengine2011}. In November 2023, a video on the Chinese Internet sparked heated discussions. A doctor diagnosed a patient while swiping the screen of his mobile phone to search. Many Internet users believed that the doctor’s behavior was inappropriate \citep{Searchengine2023}. There has also been a lot of discussion on social media platforms about whether doctors should use search engines to diagnose patients \citep{GoogleSearchengine2017}. A relevant question is ``whether doctors get mad when some patients come with their own Google research and diagnosis'' \citep{patientsSearchengine2017}. These are typical issues and dilemmas in the doctor-patient relationship. Nowadays, the rise of LLMs and their domain knowledge in healthcare and medicine may have an extra impact on the doctor-patient relationship in addition to search engines and online communities. Will patients mind if doctors use LLMs to diagnose them? This question is worth further consideration and research.

The above contents can be summarized as the ethics of ChatGPT and other large language models in public healthcare. A recent study~\citep{haltaufderheide2024ethics} found that the existing literature, which is about the ethics of ChatGPT in medicine and healthcare, mainly focuses on clinical applications and the support of health professionals and researchers. But the studies on public health are rare. 
\section{Methodology}
In this IRB-approved study, we used a combination of interviews and surveys to approach our research questions. Anyone over 18 who had used LLMs for healthcare Q\&A or information search was eligible for our study. 
% This is due to health issues that can occur in anyone. 
% They all have a need for healthcare information. 
% The inclusion criteria for interview participants are people over 18 years old who have used large language models (LLMs) for health information searches. 
The two methods are complementary to each other: surveys were administered to understand users' overall perceptions and ensure the generalizability of the findings. Specifically, through sampling statistics, we hope to understand the public's participation and attitudes. For example, what proportion of the public uses LLMs to obtain health information; what proportion of the public supports doctors using LLMs for diagnosis, etc. Interviews were used to learn about users' nuanced motivations, experiences, and concerns when using LLMs for health information searches. 
% but it is difficult to further understand more detailed information through the user’s background, personal experience, etc. 
% Interviews are just the opposite. 

\subsection{Surveys}
Between December 2023 and March 2024, we distributed the online survey on social media platforms such as X, Reedit, and Discord. In the recruitment message, we invited users who satisfied the inclusion criteria to complete our survey. The participation was voluntary, and the participants were offered a chance to enter a prize draw worth 50 dollars (for 5 participants). Since our advertising was primarily distributed through online communities, people who participated in the survey may be limited to those who were familiar with social media.

Interested participants were directed to the survey after clicking on our Qualtrics link. 
% The survey was designed using the Qualtrics platform. 
They were first shown information related to the survey; after they gave consent, they proceeded to the survey questions. The survey had a total of 23 questions (Including three short answer questions) and the estimated completion time was 10 minutes -- in the end, most participants completed the survey within an average of 5 minutes. When designing survey questions, we referred to previous research on AI healthcare chatbots \citep{nadarzynski2019acceptability} and took interview results and actual conditions into account. Specifically, the survey was divided into three parts: (1) Demographic information such as gender, ethnicity, education levels, and so on; (2) How participants searched for healthcare information and what types of healthcare information they searched for -- for example, ``Have you ever used the search engine (like Google, Bing, Baidu) to search for health information?'', ``Have you ever searched for information about medication (treatment, drugs, hospitalization)?'' (3) Participants were asked to evaluate controversial issues in the use of LLMs using a 5-point Likert scale(From ``Strongly Disagree'' to ``Strongly Agree''), e.g, ``I think it's reasonable for doctors to use LLMs to diagnose and treat conditions.'' 

During the quantitative analysis, we first removed surveys that were duplicates or looked like robot responses. When a large number of consecutive surveys appeared within a short period, we determined them as filled in by robots. We ended up receiving 219 valid surveys. Only 5 of these people had never used the Internet to search for healthcare information, so we finally used the remaining 214 responses for analysis. Similar to previous studies, we transformed most of the questions into binary variables and excluded neutral values \citep{nadarzynski2019acceptability}. For ethnicity and age, we conducted population statistics by interval. The questions involving value judgments and personal opinions were complex and did not fully conform to the binary distribution. However, they could be used to explore the acceptability of LLMs in public healthcare applications and provide important guidance for future directions.

Among the 214 participants, most participants were aged 20-29 (100) and 30-39 (86). There were 20 participants aged 40-49 years. Only 6 participants were 10-19 years old and 3 participants were 50-59 years old. There were no participants in other age groups. This shows that most of the survey participants are young people. This age distribution is similar to the age distribution of Chatgpt users\cite{statista2023}. There were 111 males and 99 females, and 4 participants did not want to say their gender or were of a third gender. There is no statistical difference in the gender ratio. The largest number of participants were white (90), followed by Asians (57) and blacks or African Americans (51), with fewer other ethnic groups. The educational level of most participants is postsecondary (188). 

\subsection{Interviews}
We mainly looked for participants through online recruitment. Overall, we used a combination of the following methods: (1) posting recruitment messages on Twitter and WeChat, and (2) snowball sampling. We briefly asked potential participants what healthcare questions they had asked LLMs. This was to confirm whether they met our inclusion criteria. In the end, we recruited 17 participants. Among them, 12 participants were recruited from WeChat and Twitter, and 5 from our contacts. 16 participants lived in the US and one in the UK. Most of them were students between 20 and 30 years old, which we acknowledged as a limitation. More demographic information can be seen in Table \ref{demographic}.

Participants were compensated 20 dollars for the interview. We planned the interviews to last less than 1 hour, and all interviews were completed within 45 minutes. Interviews were held online through Zoom. The interviews with Chinese-speaking participants were conducted in Chinese. Other participants were interviewed in English. During the interviews, one researcher asked questions and another researcher took notes. We recorded the interviews and saved transcripts of the conversations, which were afforded by Zoom, upon participants' permission. Participation in the interview was completely voluntary, and participants could exit at any time. 

The interviews were semi-structured. We started by asking about the basic background of the participants. We divided the interview questions into three parts: (1) motivations for using LLMs for healthcare information; (2) misinformation and privacy issues encountered when using LLMs; and (3) the role of LLMs in the doctor-patient relationship. A detailed list of questions is provided in the appendix. 

After each interview, we held a brief meeting to discuss emerging themes. Two authors independently conducted open and axial coding of the interview transcripts \citep{strauss1990qualitative}, and regularly met to compare their results and discuss the coding process. If the two coders could not reach a consensus, a third researcher would join the discussion to help make a final decision. 
% Emerging themes included ..., ..., ..., ..., and so on. 
In the paper, we use anonymized quotes to illustrate the main themes.

\begin{table*}
\centering
\begin{tabular}{lllllll}
\hline
ID  & Gender & Age   & Educational Level & Race & Profession         \\ \hline
P1  & F      & 22    & Bachelor & Asian        & Student  \\
P2  & F      & 25    & Master & Asian      & Student                    \\
P3  & F      & 22 & Bachelor & Asian        & Student             \\
P4  & F      & 23 & Bachelor & Asian      & Student            \\
P5  & F      & 23    & Bachelor   & Asian         & Student                    \\
P6  & F      & 23   & Bachelor   & Asian         & Student                     \\
P7  & M      & 26    & Bachelor & Asian      & Software Engineer           \\
P8  & M      & 27    & High School   & Black      & Freelancer                  \\
P9  & M      & 28    & Bachelor   & Black      & Freelancer                     \\
P10 & F      & 23    & Bachelor & Black      & Salesperson            \\
P11 & M       & 27    & Master & Black    & Economic Researcher           \\
P12 & M        & 21    & Bachelor & Black      &Student     \\
P13 & M        & 24		& Bachelor & White     & Student          \\ 
P14 & M        & 29		& Bachelor & White      & Freelancer          \\ 
P15 & M        & 23		& Bachelor & Black      & Student         \\
P16 & F        & 23		& Bachelor & Asian      & Student        \\
P17 & M        & 22		& Bachelor & Asian      & Student        \\  \hline
\end{tabular}
\caption{Demographic information of our participants. }~\label{demographic}
\end{table*}
\section{Results}
In this section, we answer our research questions. As explained in the methodology, both qualitative and quantitative methods were used to approach the research questions. Each method has its advantages and disadvantages and can complement each other. Therefore, research questions and research methods do not correspond one-to-one; a research question may involve one method or both.

\subsection{How The Public Obtain Healthcare Information? (RQ1)}
LLMs are gaining popularity as a means of healthcare information seeking. Among the 219 valid surveys, 214 participants had used the Internet (including but not limited to search engines, online communities, AI-based healthcare chatbots, and LLMs) to search for healthcare information.
% Therefore, we can conclude that the vast majority of the public has used search engines to obtain health information. 
208 (97\%) participants had searched for healthcare information from search engines, 178 (83\%) participants had searched for healthcare information from online communities, and 166 (78\%) participants had obtained healthcare information from LLMs. This shows that search engines are still the most mainstream way to obtain healthcare information. LLMs like ChatGPT have only been around for about a year, but they have already become a frequently used way for the public to obtain health information. In contrast, only 91 (43\%) participants had used Al-led healthcare chatbots, which have existed for a longer time than LLMs. The percentage of users who used chatbots for healthcare Q\&A was slightly larger than in a previous study~\citep{nadarzynski2019acceptability}, but chatbots were still not a mainstream way for the public to obtain healthcare information. We confirmed this during the interviews. 

There was more than one way to obtain healthcare information from the Internet, and they were often used together for better information outcomes. Of the 166 participants who had used LLMs to obtain health information, 144 participants cross-validated the healthcare information obtained from LLMs with healthcare information obtained from other sources. This suggests that LLMs have a complementary relationship with other ways of obtaining healthcare information in the digital niche. During the interviews, we found that due to the popularity of online consultation platforms in China \citep{ding2020getting, hao2015development, hao2017tale}, Some participants (P4, P6) originally from China had heavily drawn on them for healthcare information. The students majoring in medicine (P3, P16) would also use more professional medical databases, such as the Merck Manual of Diagnosis and Therapy. Other findings were largely consistent with the survey -- the participants interviewed had all used search engines to search for healthcare information before, and most had obtained health information through online communities. Among the 17 participants who had used LLMs to search for health information, only P5 and P9 had used AI-led chatbots, this also shows that the popularity of AI-led Chatbots is not high. 

In the survey, based on some previous studies~\citep{rice2006influences,nadarzynski2019acceptability, xiao2014factors}, we roughly divided the healthcare information obtained into four categories: (1) daily life health information (such as reasonable diet); (2) medication information (such as drugs and hospitalization); (3) symptom self-diagnosis (such as searching for disease); and (4) searching for health information again after being diagnosed by a doctor due to doubts about the treatment plan. Among the 214 participants who had searched for health information, 194 (91\%), 197 (92\%), 190 (89\%), and 176 participants had searched for the above four types of information respectively. The search frequency for the four types of information was all very high and there was no significant difference. This phenomenon showed that the public was not only paying attention to diseases but also their healthcare in daily life. Similarly, during the interviews, we found that the participants mostly used ChatGPT to search for symptom-related information, but also used it to search for daily healthcare information such as reasonable diets and fitness plans. For example, P2 only uses Chatgpt to initially develop a fitness plan:

\begin{quote}
    \textit{``Whenever I don't know what to do this week, I always ask Chatgpt first and let it make a preliminary fitness plan for me. Although it may be superficial, it can give me an idea.''}
\end{quote} 

In terms of frequency of searching for health information, 121 participants in the survey searched for health information once a month or more often; during the interview, most participants said they would search for health information every one to two weeks. A few interview participants like P1 searched for healthcare information nearly every day.

\subsection{Motivation for Using LLMs for Healthcare Information (RQ2)}
During the interviews, we asked the participants in detail about their motivations for using LLMs to obtain health information. We were particularly curious about why participants chose to use LLMs like ChatGPT to obtain healthcare information rather than other methods such as search engines, online communities, or consulting doctors online and offline.

Compared with search engines, many interview participants used ChatGPT to obtain healthcare information because LLMs are faster and more accurate. Most of the participants were students. Before using ChatGPT to search for health information, they had used it to search for solutions to study problems. They believed that ChatGPT was more helpful to them than search engines in solving academic problems. This prompted them to start using ChatGPT for healthcare information as well. 

Firstly, the density of healthcare information obtained by using search engines was low, because users needed to open many websites one by one to browse information, which took a lot of time. On the contrary, ChatGPT provided targeted and organized information in a short time. For users, using a search engine was not as efficient as using ChatGPT. P14 said:

\begin{quote}
    \textit{``When I use Google, I have to carefully judge whether each search result is relevant to the information I want to know. However, Chatgpt was able to give me what Iwant to know very quickly.''}
\end{quote}

Secondly, the quality of healthcare information obtained from search engines varied. The information was mostly accessed from social media or online communities, where a lot of misinformation existed. Relatively speaking, participants tended to think that the health information given by ChatGPT was more reliable due to the trust built through previous Q\&A sessions irrelevant to healthcare. During the interviews, only two participants stated that they had used AI-based healthcare chatbots before using ChatGPT. This suggested that the participants' trust in LLMs for obtaining health information came from their good performance in providing information in other fields.

In the comparison between LLMs and online communities, most participants admitted that the purpose of using online communities was not only to obtain healthcare information. Online communities also helped them communicate and connect with other patients and encourage each other. P1 elaborated on this point:
\begin{quote}
    \textit{``I can get some objective information using search engines or ChatGPT. But the mutual encouragement between patients and the subjective feelings about the use of different drugs can only be obtained by communicating with real people in the community.''}
\end{quote}
P5 also had the same idea. She believed that some subjective feelings of drug use could not be described in detail by Chatgpt. ChatGPT rarely provided emotional support.  In other words, users have different orientations toward obtaining health information from LLMs and online communities. 

The participants thought obtaining healthcare information from LLMs was more convenient and quick than offline or online medical consultations with doctors. Making an appointment with a doctor offline not only costs money but also takes a lot of time; even if one made an appointment with a doctor online, there was no guarantee that the doctor would always be online to answer questions. However, all the participants acknowledged that if they felt their condition was serious and needed medical equipment for examination, they would still choose to go to the hospital. In other words, all participants would only use LLMs when they were experiencing minor discomfort or searching for daily healthcare information.

From the motivations of the interview participants, we can further infer the niche of LLMs in healthcare information seeking. In general, although Chatgpt is a "chatting" robot, most users who use Chatgpt to search for medical and health information, regard Chatgpt (or other LLMs) as a supplement or substitute for search engines. In other words, LLMs have competitive and complementary relationships with search engines and have a complementary relationship with online consultation and online communities. Users expect LLMs and search engines to provide accurate medical advice; expect online communities to provide mental support and expect online medical consultations to provide professional medical services across spatial limitations. 

\subsection{Public Concerns and Choices in Using LLMs for Healthcare (RQ3)}
In this section, we mainly introduce the public’s concerns and choices about the two major issues existing in the practical application of LLMs for healthcare information: (1) misinformation and accuracy of responses; and (2) doctor-patient relationship issues. In addition, we will briefly touch on privacy concerns.

\subsubsection{Misinformation and Inaccuracy of LLM Responses}
LLMs were deemed more accurate than other information-seeking channels for healthcare information. In the survey, we designed two Likert-scale questions to probe this issue: ``You can find more accurate health information with LLMs (like ChatGPT) than search engines or online communities'' and ``Compared with using search engines or online communities, using LLMs can reduce misinformation.'' For the first statement, 109 (51\%) participants ``agree'' or ``strongly agree'' with it, and 36 (17\%)  participants ``disagree'' or ``strongly disagree'' with it. For the second statement, 111 (52\%) participants ``agree'' or ``strongly agree'' with it, and 51 (24\%) participants ``disagree'' or ``strongly disagree'' with it. As we can see, more than half of the participants agreed that LLMs are more accurate and can produce less misinformation than other methods like search engines. 

To obtain a more accurate picture of what the public thought, we interviewed participants about this issue as well. Most participants stated that LLMs provided more accurate health information than search engines or online communities. P10 was one of them,
\begin{quote}
    \textit{``Search engines can actually easily lead me to some other online community, or just serve me some ads... And sometimes it's hard for me to judge the accuracy of the information given by these linked pages. But ChatGPT can give me some very intuitive information. Although such information may be very superficial, I think they are correct.''}
\end{quote}

A few participants (P3, P5, and P16) whose majors or jobs were related to medicine and biology held different views. For example, P3’s major was animal medicine. She thought ChatGPT was not up-to-date when it came to medical knowledge:
\begin{quote}
    \textit{``When I was looking for information about coursework, I entered my question into ChatGPT, but it gave me a confusing answer. I then went straight to MSD Manual and confirmed that the answer was incorrect. I don’t think ChatGPT has done much to reduce the misinformation associated with some of the latest medical research advances.''}
\end{quote}
These participants were medical students and practitioners and thus had a higher level of medical knowledge than the general public. They were able to effectively detect misinformation that appeared when ChatGPT answered healthcare questions. At the same time, they acknowledged that misinformation was often limited to the latest medical research; for health information that the public searched for in their daily lives, ChatGPT was still able to give superficial/ambiguous yet relatively accurate/correct answers. 

During the interviews, we also asked participants how they judged misinformation, or how they prevented ``hallucinations'' from LLMs. Most of the participants did not completely give up using search engines or online communities, and they would cross-verify the health information obtained from LLMs with health information obtained through other channels. Some participants, such as P12 and P14, would consult doctors after asking LLMs for health information, and their trust in LLMs was enhanced by the doctor's feedback. However, there are exceptions. P8 said that since using Chatgpt, he has never used a search engine again because he hates the misinformation on Google.

\subsubsection{Doctor-Patient Relationship}
In the survey, we designed three questions to explore this question: ``Patients using LLMs to self-diagnose before asking the doctor can improve the doctor's diagnostic performance,'' ``I think doctors using LLMs to diagnose and treat conditions is reasonable'' and ``I think doctors using LLMs for medical assistance work (such as writing medical records) is reasonable.'' For the first statement, 120 (56\%) participants ``agree'' or ``strongly agree'' with the statement, and 42 (20\%) participants ``disagree'' or ``strongly disagree'' with the statement. For the second statement, 106 (50\%) participants ``agree'' or ``strongly agree'' with the statement, and 47 (22\%) participants ``disagree'' or ``strongly disagree'' with the statement. For the last statement, 140 (65\%) participants ``agree'' or ``strongly agree'' with the statement, and 30 (14\%) participants ``disagree'' or ``strongly disagree'' with the statement. It can be seen that most people thought it reasonable for doctors to use LLMs to perform auxiliary work such as writing medical records. When it came to using LLMs for diagnosis or self-diagnosis, the participants were more doubtful. Most participants also supported that the use of LLMs for self-diagnosis can help doctors make better diagnoses. 

During the interviews, most of our participants believed that self-diagnosis assisted by LLMs could help doctors better judge their conditions. This is consistent with the survey. P6 talked about the importance of knowing one's own medical history and physical conditions before seeing a doctor:
\begin{quote}
    \textit{``Doctors are also human beings, not gods. If you don't understand your past medical history and your physical condition, doctors are also likely to make wrong judgments. Therefore, I believe that self-diagnosis in advance is also an important step to help doctors diagnose.''}
\end{quote}

Most participants did not accept doctors to use LLMs for diagnosis. P4 thought relying on LLMs to diagnose common diseases indicated unprofessionalism of the doctors, and only approved of doctors to diagnose rare diseases with the help of LLMs:
\begin{quote}
    \textit{``I think the diagnosis of common diseases should be something that a doctor should have mastered. If ChatGPT is required for this type of disease, I will think this doctor is very unprofessional. But for rare diseases, I think doctors are inherently inexperienced and it is reasonable to resort to other methods.''}
\end{quote}
P17 thought doctors should apply their own judgment and thinking after LLMs gave a diagnosis:
\begin{quote}
    \textit{``I can accept the doctor using LLMs to make a diagnosis, but after the search results are found, the doctor must tell me why he thinks the diagnosis is correct.''}
\end{quote}

However, a few participants believed that LLMs could be used for diagnosis since AI was an unstoppable trend and would eventually replace doctors. P7 noted:
\begin{quote}
    \textit{``Artificial intelligence is a need for the development of medicine. It is impossible for doctors to master all medical knowledge. Using ChatGPT can improve the efficiency of diagnosis. I believe that there will be artificial intelligence that can replace doctors in the future, so I now support doctors using ChatGPT for diagnosis.''}
\end{quote}

As for auxiliary work such as writing medical records, almost all interview participants thought it was reasonable and LLMs helped doctors improve their productivity. 
Only P5 had a different view. She said:

\begin{quote}
    \textit{``Although I can understand doctors using Chatgpt to write medical records, I don't want them to do it in front of me because it makes me feel that the doctor is not professional.''}
\end{quote}

This shows that some members of the public may rationally support LLMs entering the healthcare industry, but they are still unable to accept it emotionally.
\section{Discussion}
In this section, we discuss recommendations for the public to use LLMs to obtain health information and conduct self-diagnosis and recommendations for doctors to use LLMs to diagnose patients. We further give recommendations for future research in medical LLMs and give inventions for the future development direction of medical LLMs.

\subsection{Recommendations for The Public}
Since everyone's physical condition is different, the same symptoms may come from different diseases for different people. We ought to fully consider these medical backgrounds and make contextual recommendations for the public regarding using LLMs to obtain health information to maximize public interests.

According to our results, although LLMs cannot completely solve the problem of misinformation generated by search engines and online communities, they have advantages from a development perspective -- on the one hand, LLMs do provide more accurate healthcare information; on the other hand, LLMs can help users get the health information they need in a shorter time and a concise manner. As such, using LLMs to obtain health information is an effective complement to other information channels. The advantages of LLMs stand out particularly when the public wants to obtain daily information or when they are mildly unwell -- health Q\&A afforded by LLMs can already help them cope with most health conditions in life. 
% Medical treatment is still needed at this time. 
But we also need to point out, LLMs only provide relatively simple and error-free answers. For serious and complex diseases, or when users need to develop personalized fitness/weight loss programs for themselves, LLMs are powerless. When it comes to more serious symptoms or illnesses, a thorough examination by doctors is still necessary. Of particular concern are "hallucinations": some patients believe that they do not have the disease, so they constantly enter "prompts" denying the existence of the disease to Chatgpt. Chatgpt will also output content that patients want to see "there is no disease" following the user's ideas. Patients may therefore delay treatment~\citep{saenger2024delayed}. Similarly, LLMs still cannot be personalized perfectly today. Due to individual differences, if users want to develop professional health plans, such as fitness plans or diet plans based on their conditions, they still need to seek advice from professionals.

\subsection{LLMs and The Doctor-Patient Relationship}
Many survey participants did not agree that self-diagnosis could help doctors better judge their conditions. However, during the interviews, almost all participants admitted that after self-diagnosis, they could better communicate with doctors and enable doctors to better understand their physical conditions. We believe that everyone is the first person responsible for their own health~\cite {resnik2007responsibility}. If one has a good understanding of their own health status and living habits and can express them to doctors more accurately, the doctors can better judge the patient's conditions. Therefore, LLMs' role in helping patients conduct preliminary self-diagnosis before seeing a doctor is promising -- this is a sign of being responsible for one's health.

The issue of whether doctors should use LLMs to diagnose patients has generated significant controversy, both in surveys and in interviews. On the one hand, using LLMs for diagnosis made doctors appear unprofessional, according to the participants. On the other hand, doctors have their limitations and cannot make accurate judgments at all times. For example, medical resources are limited and doctors in less socioeconomically developed areas may not have sufficient professional knowledge and expertise \citep{ding2020getting,haltaufderheide2024ethics}. It is not uncommon for doctors to be physically harmed or even killed due to unsatisfactory treatment outcomes \citep{he2016explaining}. There are also some doctors spreading misinformation during the COVID-19 epidemic~\cite {sule2023communication}. Therefore, search engines or LLMs may be an important aid for doctors to make diagnoses or treatment plans. 
% We agree with what P15 said, i.e., 
As several participants suggested, doctors can use LLMs for diagnosis, but they need to first communicate with the patients about their use and obtain their agreement. Further, doctors should apply their judgment to LLM-generated output and explain their thought processes to the patients to build trust and ensure reliable medical outcomes. 
% the basis for the judgment also needs to be explained to the patient. 
% This can not only satisfy patients but also improve the accuracy of diagnosis.

Most participants approved doctors of using LLMs to do auxiliary work such as writing medical records. Such LLM use reduces the workload of doctors and improves their work efficiency. Again, patients want to reserve the right to know how LLM is used in the process; if LLMs generate responses that contain errors, doctors should correct them in time to avoid mistakes. As we can see, transparency and accountability \cite{kilhoffer2023ai} are important factors in LLM-assisted medical practices. Future research should consider documenting ethical incidents and errors when using LLMs for healthcare to help in this regard \cite{wei2022ai}.

\subsection{The Future of Public Health and Trusted Healthcare LLMs}
From the perspective of digital niche, most users expect LLMs to be more than just a chatbot when seeking healthcare information, but a tool that replaces search engines. Users expect to obtain healthcare information accurately and timely through LLMs, which means that people have similar expectations of LLMs as professional doctors. If LLMs are expected to do the jobs of a doctor, the first thing we need to discuss is what scientific and ethical principles a trustworthy doctor should abide by. Among these, the public is particularly concerned about medical misinformation. This has triggered a scientific and ethical debate: Is the information provided by doctors "misinformation"? How can we identify "misinformation" in medical treatment? Medicine and Healthcare are rapidly developing -- academic results that are correct today may be overturned or proven to have certain limitations tomorrow~\citep{singhal2023large}. The medical and healthcare field is a rapidly developing field, and many cutting-edge studies may contain errors or even academic fraud, such as cardiac stem cells \citep{heartstemcell2018}. So we must have a clear definition of misinformation in medicine and healthcare, otherwise there will be confusion. In June 2022, in order to curb the spread of misinformation, the American Medical Association adopted a policy outlining specific steps to identify and curb false and misleading medical information~\citep{AMS2022}. Based on this guidance, we provide some inspiration for future trustworthy medical LLMs.

Timely and rapid updating of the database (corpus) is critical. In the guidance, medical misinformation is defined as: "Health-related information or claims that are false, inaccurate or misleading, according to the best available scientific evidence at the time". We cannot ask doctors to answer medical questions that cannot be explained by current evidence, nor can we ask doctors to diagnose major diseases without any mistakes. The same is true for LLMs. However, our scientific research needs that LLMs can always follow the latest scientific research progress. Today, with the rapid progress of scientific research, only by constantly updating the database can LLMs ensure that they can provide users with the most accurate information based on current scientific evidence. 

Effective methods of identifying misinformation are needed. For the field of medicine and health, in addition to misinformation on social media or generated by other LLMs, a more unique challenge is identifying misinformation in the literature. If LLMs are required to provide information that is most consistent with current scientific evidence, it is inevitable to incorporate the latest research literature into the database. However, a survey showed that of the more than 5,000 neuroscience articles published in 2020, 34\% may be plagiarized or made up. In medicine, this proportion is 24\%~\citep{Science2023}. Prior research pointed out that ChatGPT may cite false medical literature \citep{frosolini2023reference} In an environment where LLMs have shifted from model-driven to data-driven~\citep{he2023survey}, the quality of data determines the accuracy of the model. Filtering out harmful articles is of great significance to improve the accuracy of LLMs.

Balancing autonomy with medical advice. Doctors indeed need to abide by the principle of autonomy and respect patients' right to choose their medical solutions, but this does not mean that doctors should excessively cater to patients' ideas and make patients choose wrong or outdated diagnoses and treatment options. A good doctor should also be able to persuade patients to accept the correct diagnosis and treatment plan. This puts forward higher requirements for the dialogue system of LLMs. While avoiding hallucinations, LLMs need to use humane methods and language to help and even persuade patients to choose the best treatment options.

In the field of daily health, such as weight loss and fitness, personalized health care LLMs are a direction worthy of attention. Current LLMs still fail to develop personalized solutions based on user needs ~\citep{yang2024social}. The ultimate goal of disease diagnosis and treatment is to restore patients to health, but users who use LLMs for daily health activities are likely to have different needs. Thus, understanding user needs and designing personalized LLMs become increasingly important.

Combining LLMs with search engines. Although our research found that LLMs and search engines compete, this does not mean that the two do not have a complementary relationship. A study has shown that users prefer to use search engines to search for direct, fact-based queries, and use LLMs to search for complex natural language processing problems \cite{caramancion2024large}. Many works combine LLMs with search engines and point out that LLMs can be excellent search engines \cite{ziems2023large,vu2023freshllms}. This shows that integrating the advantages of both to facilitate users to search more conveniently will be a future trend.

\subsection{Limitation}
There are several limitations of our research. Firstly, limited by information channels and inclusion criteria, almost all our interview participants were between the ages of 20 and 30, and lots of them were students. Most of the survey participants are between the ages of 20 and 39, although this is roughly consistent with the age group of Chatgpt users~\cite {statista2023}. Our interview sample was also limited to Asian and Black people. As a result, we were unable to learn about the experiences and challenges of other age groups and races when using LLMs for healthcare. For example, Middle-aged and elderly people may have higher and more complex needs for healthcare. We hope future research should consider more diverse demographic groups, especially the middle-aged and elderly. 
% but considering our interview inclusion criteria, we are temporarily unable to know whether middle-aged and elderly people will use LLMs for health information search. 

Secondly, we only conducted a simple qualitative analysis of the survey data, without conducting a rigorous quantitative analysis and taking into account the correlation between different issues. For example, does gender affect people's acceptance of doctors using LLMs for diagnosis? In future work, we will further collect data and use quantitative analysis methods to study the public's use of LLMs for health information searches. 

Thirdly, our team does not include members who are professionally engaged in medical research. In fact, during surveys and interviews, we also collected questions entered by users into Chatgpt. Future research could consider probing how people use LLMs to cope with specific medical conditions from a professional perspective. 
% In the future, we may seek opinions from professional medical personnel and further promote our research.
\section{Conclusion}
We use a mix of surveys and interviews to study the concerns and choices of the public when using large language models for healthcare. Compared with prior research on public healthcare and large language models, our research has the following three contributions: first, through the surveys, we learned about the different ways the public uses to obtain healthcare information from LLMs; second, through both surveys and interviews, we learned about different motivations of the public for using LLMs for healthcare; third, we learned about the public's value judgments when facing some of the controversial issues, such as misinformation and doctor-patient relationships. We concluded by providing recommendations for the public to use LLMs better to search for health information and for doctors to use LLMs for diagnosis more ethically. We also pointed out future research directions in LLMs for public health.

\bibliographystyle{ACM-Reference-Format}
\bibliography{acmart}

\section{Appendix}
\subsection{Interview Questions}
\label{interview}

(1) Briefly describe your age, job, education, race, and religion.

(2) What LLMs did you use (ChatGPT 3.5 or 4)?

(3) Have you heard of (or used) health chatbots before LLMs?

(4) Can you give a specific example of using LLMs to find health information (healthcare Q\&A/judging diseases when you are unwell/seeking medical treatment when you are sick)

(5) How did you obtain healthcare information before using LLM (TV/Internet/magazines/online communities)?

(6) Why did you start using LLMs?

(7) Compared with the information acquisition methods in the past, what do you think are the advantages and disadvantages of LLMs in obtaining information? Can you give some examples?

(8) Frequency of using the Internet or LLMs to search for health information?

(9) Have you found mistakes or misinformation when you asked LLMs for healthcare information? Did they happen very often? Could you give some specific examples? (What questions were asked of the LLM and what were the responses of the LLM?)

(10) Did you have similar problems with other ways of information seeking you used before LLMs came along? 

(11) Who do you think should be held accountable in the event of serious consequences (company, individual, or other parties))?

(12) How do you judge the authenticity of health information generated by LLMs? For example, do you cross-validate the information you get?

(13) Compared with consulting a doctor (or fitness coach), what do you think are the advantages of using LLMs? (Why would you rather go to LLMs than go to a doctor?)

(14) If you go to see a doctor after using LLMs, is the judgment of LLMs and the doctor always consistent? If not, who do you think gives more accurate advice? 

(15) Do you think using LLMs before consulting a doctor can help the doctor better judge your physical condition? Why?

(16) What are your thoughts on doctors using LLMs to diagnose you? (If doctors use LLM only for auxiliary work, such as writing medical records, can you accept it? 

(17) Do you have any other concerns in addition to what we mentioned above?

(18) Can you please send us a list of questions you entered into ChatGPT? We later want to re-enter these questions into ChatGPT to see if there are hallucinations and misinformation.

\newpage

\subsection{Survey Questions}
(1) Your age.

(2) Your gender.

(3) Your ethnicity.

(4) Your education Level.

(5) Have you used large language models (like Chatgpt)?

(6) Health information search frequency (Either use LLMs, Internet, TV or magazines).

(7) Have you ever used the search engine(like Google, Bing, Baidu) to search for health information?

(8) Have you ever used the online community (like Reddit, X(Formerly Twitter)) to search for health information?

(9) Have you ever used the LLMs (Like ChatGPT) to search for health information?

(10) Have you fact-checked information obtained from LLMs (via search engines, online communities or other methods).

(11) Have you used of health chatbots (Ada, K Health, etc) before LLMs?

(12) Have you ever searched general health information (such as lose weight, fitness, Eat properly)?

(13) Have you ever searched information about medication (treatment, drugs, hospitalization)?

(14) Have you ever searched information about symptoms (such as headache and stomacher)?

(15) Have you ever searched information about diseases and treatment options after diagnosis (such as COVID-19)?

(16) You can find more accurate health information with LLMs (like ChatGPT) than search engines or online communities. (Agree or Disagree)

(17) Compared with using search engines or online communities, using LLMs can reduce misinformation. (Agree or Disagree)

(18) Patients using LLMs to self-diagnose before asking the doctor can improve the doctor's diagnostic performance. (Agree or Disagree)

(19) I think doctors using LLMs to diagnose and treat conditions is reasonable. (Agree or Disagree)

(20) I think doctors using LLMs for medical assistance work (such as writing medical records) is reasonable. (Agree or Disagree)

\newpage

% \begin{table*}
% \centering
\begin{longtable}{l|l}
\hline
Variable & Total of the sample(\%)                                             \\ \hline
Total Responds & 214 (100\%) \\
Age &                                  \\
\quad \quad 20-29 & 100 (47\%)                      \\
\quad \quad 30-39 & 86 (40\%)                      \\
\quad \quad 40-49 & 20 (9\%)                      \\
Gender &  \\
\quad \quad Male & 111 (52\%)                      \\
\quad \quad Female & 99 (46\%)                      \\
Ethnicity & \\
\quad \quad White & 90 (42\%)                      \\
\quad \quad Asian & 57 (27\%)                      \\
\quad \quad Black & 51 (24\%)                        \\
Education Level & \\
\quad \quad Under Postsecondary & 26 (12\%)                      \\
\quad \quad Postsecondary & 188 (88\%)                      \\
Past LLMs Use & \\
\quad \quad Yes & 195 (91\%)                      \\
\quad \quad No & 19 (9\%)                      \\
Health Information Seeking & \\
\quad \quad Several times per year & 93 (43\%)                      \\
\quad \quad Every month or more often & 121 (57\%)                      \\
Past Use Search Engines for Healthcare Information & \\
\quad \quad Yes & 208 (97\%)                      \\
\quad \quad No & 6 (3\%)                      \\
Past Use Online Community for Healthcare Information & \\
\quad \quad Yes & 178 (83\%)                      \\
\quad \quad No & 36 (17\%)                      \\
Past Use LLMs for Healthcare Information & \\
\quad \quad Yes & 166 (78\%)                      \\
\quad \quad No & 48 (22\%)                      \\
Fact-checked Information Obtained from LLMs & \\
\quad \quad Yes & 144 (67\%)                      \\
\quad \quad No & 22 (10\%)                      \\
Past Use Health Chatbots (Like Ada) for Healthcare Information & \\
\quad \quad Yes & 91 (43\%)                      \\
\quad \quad No & 123 (57\%)   \\
To Seek General Health Information & \\
\quad \quad Yes & 194 (91\%)                      \\
\quad \quad No & 20 (9\%)   \\
To Seek Medication & \\
\quad \quad Yes & 197 (92\%)                      \\
\quad \quad No & 17 (8\%)   \\
To Seek Symptoms & \\
\quad \quad Yes & 190 (89\%)                      \\
\quad \quad No & 24(11\%)   \\
To Seek Diseases and Treatment Options after Diagnosis & \\
\quad \quad Yes & 176 (82\%)                      \\
\quad \quad No & 38 (18\%)   \\
LLMs can Find More Accurate Health Information than Search Engines or Online Communities\\
\quad \quad Agree and Somewhat Agree & 109 (51\%)                      \\
\quad \quad Disagree and Somewhat Disagree  & 36 (17\%)   \\
LLMs can Reduce Healthcare Misinformation than Search Engines or Online Communities\\
\quad \quad Agree and Somewhat Agree & 111 (52\%)                      \\
\quad \quad Disagree and Somewhat Disagree  & 51 (24\%)   \\
Using LLMs to Self-diagnose can Improve Doctors' Diagnostic Performance \\
\quad \quad Agree and Somewhat Agree & 120 (56\%)                      \\
\quad \quad Disagree and Somewhat Disagree  & 42 (20\%)   \\
Doctors Use LLMs to Diagnose and Treat Conditions is Reasonable \\
\quad \quad Agree and Somewhat Agree & 106 (50\%)                      \\
\quad \quad Disagree and Somewhat Disagree  & 47 (22\%)   \\
Doctors Use LLMs to Diagnose and Treat Conditions is Reasonable \\
\quad \quad Agree and Somewhat Agree & 140 (65\%)                      \\
\quad \quad Disagree and Somewhat Disagree  & 30 (14\%)   \\

\caption{Demographic information of our participants in survey. }~\label{demographic}
\end{longtable}
% \caption{Demographic information of our participants in survey. }~\label{demographic}
% \end{table*}

%\end{sloppypar}
\end{document}